# A microscopic cranking model and its connection to other collective models for uni-axial rotation

P. Gulshani

NUTECH Services, 3313 Fenwick Crescent, Mississauga, Ontario, Canada L5L 5N1
Tel. #: 647-975-8233; matlap@bell.net

A microscopic time-reversal invariant cranking model (*MCRM*) for nuclear collective rotation about a single axis and its coupling to intrinsic motion is derived. The *MCRM* is derived by transforming the stationary nuclear Schrodinger equation using a collective rotation-intrinsic product wavefunction, imposing no constraints on either the wavefunction or the space-fixed nucleon coordinates, and using no relative co-ordinates. Therefore, this formulation collective and intrinsic motions are described by the same space-fixed co-ordinates and momenta and within the same phase space. The derivatives of the collective-rotation angle are defined in terms of a combination of rigid and irrotational collective flows of the nucleons. The collective wavefunction is chosen to be an eigenstate of the angular momentum, yielding a *MCRM* Schrodinger equation for the intrinsic wavefunction that contains a cranking Coriolis energy term that is linear in the angular momentum and shear operators, a collective centrifugal energy term, and a rotation-fluctuation energy term. In absence of the irrotational-flow component and fluctuation energy term, the *MCRM* equation reduces to that of the conventional cranking model (*CCRM*), but with a dynamic rigid-flow angular velocity and rigid-flow centrifugal-energy term. The expectation of the angular momentum operator, which is the sum of the collective rotation angular momentum and the expectation of the angular momentum in the intrinsic state, would reduce to that in the *CCRM* if the collective rotation angular momentum were small. However, it is shown that, even for the simple case of the anisotropic harmonic oscillator mean-field potential in $^{20}_{10}Ne$, the collective rotation angular momentum is not small in the current version of the *MCRM*, and that this problem needs further study. It is also shown that the *MCRM* Schrodinger equation is reducible to the equations of the particle-plus-rotor, phenomenological and microscopic collective rotation-vibration, and two-fluid semi-classical collective models.

*PACS number: 21.60.Ev, 21.60.Fw, 21.60.Jz*

*Keywords:* canonical transformation; multi-particle rigid and irrotational flow rotation; microscopic and conventional cranking and other collective models; dynamic angular velocity; time-reversal invariance

## 1. Introduction

The self-consistent conventional cranking model (*CCRM*) [1,2] is frequently used to study collective rotational properties of deformed nuclei [3-27 and references therein]. It has been proven to be a relatively simple, transparent, and successful method for investigating collective rotational properties and phenomena in deformed nuclei[1]. The model assumes that the anisotropic nuclear potential $V$ is rotating at a constant angular frequency $\omega_{cr}$ about $x$ or 1 axis. The model time-dependent Schrodinger equation[2]:

$$i\hbar \frac{\partial}{\partial t}|\Psi_{cr}\rangle = H_{cr}|\Psi_{cr}\rangle \qquad (1)$$

where:

$$H_{cr} \equiv \frac{1}{2M}\sum_{n,j=1}^{A,3} p_{nj}^2 + V_{cr}(\vec{r}_n), \qquad \vec{r}_n = R(\omega_{cr}t)\vec{r}_n' \qquad (2)$$

and $R$ is an orthogonal matrix and $\vec{r}_n'$ is the n$^{th}$ particle coordinate relative to the rotating frame, is then unitarily transformed to the rotating frame:

$$|\Psi_{cr}\rangle = e^{-i(\omega_{cr} L + E) t/\hbar}|\Phi_{cr}\rangle \qquad (3)$$

One then obtains the stationary *CCRM* equation[3]:

$$H_{cr} \cdot \bar{\Phi}_{cr} \equiv (H - \omega_{cr} \cdot L) \cdot \bar{\Phi}_{cr} = \bar{E}_{cr}\bar{\Phi}_{cr} \qquad (4)$$

where $L$ is the total angular momentum operator. The angular velocity $\omega_{cr}$ is then determined by requiring the expectation of $L$ to have a fixed value $\hbar J$:

$$\hbar J \equiv \langle \Phi_{cr}|L|\Phi_{cr}\rangle \qquad (5)$$

The energy $E_{cr}$ in a space-fixed frame is then given by:

$$E_{cr} = \langle \Phi_{cr}|H|\Phi_{cr}\rangle = \langle \Phi_{cr}|(H_{cr} + \omega_{cr} \cdot L)|\Phi_{cr}\rangle = \bar{E}_{cr} + \omega_{cr} \cdot \langle \Phi_{cr}|L|\Phi_{cr}\rangle \qquad (6)$$

The effective dynamical moment of inertia $\mathcal{I}_{eff}$ is not an observable and must be deduced from other predicted and measured nuclear properties. A definition of $\mathcal{I}_{eff}$, which is adopted from a rigid-body rotation and is commonly used, is given at each value $J$ by the excitation energy $\Delta E_J$:

---

[1] Of-course, there are many other models that have had various degrees of success in predicting collective nuclear properties. These other models are not discussed in this article because this article is concerned only with a microscopic derivation of the *CCRM*.

[2] Clearly, this time-dependent description of the rotational motion is classical in nature because the c-number parameter $\omega_{cr}$ is not an operator acting on a nucleon probability distribution.

[3] Eq. (4) can also be derived from a variation of the Schrodinger equation subject to energy minimization, with the wavefunction $\Phi_{cr}$ constrained to give a fixed value for the expectation of the angular momentum operator.

$$\frac{2\mathcal{J}_{eff}}{\hbar^2} = \frac{4J-2}{\Delta E_J - \Delta E_{J-2}} \quad (MeV)^{-1} \tag{7}$$

$$\Delta E_J \equiv E_J - E_0 \tag{8}$$

Because $\omega_{cr}$ is a *c*-number and hence the rotation in the *CCRM* is externally driven, the model is semi-classical and phenomenological in nature, and Eq. (4) is time-reversal non-invariant. It is therefore desirable to have a cranking model where the rotation is driven by the motions of the nucleons instead of externally, i.e., it is desirable to derive the model microscopically, as suggested in [2,4,5,7,12,18,28]. In several studies starting from first principles, Eq. (4) was derived, with various degrees of success, using canonical transformations, angular momentum projection, generator-coordinate, and equation of motion or generalized density matrix methods, and using various approximations and assumptions such as redundant coordinates, large deformations, expansion in power of the angular momentum, and truncating density matrix expansion equations at first order, etc. [4,5,12,18,29-34].

In this article, we derive, simply and from first principles, a microscopic time-reversal invariant cranking type model (*MCRM*) that can be readily reduced to the *CCRM* in Eqs. (4) and (6). The *MCRM* is derived by transforming the stationary nuclear Schrodinger equation using a collective rotation-intrinsic product wavefunction[4], where the collective rotation angle is defined in terms of a combination of rigid and irrotational collective motions of the nucleons. No constraints are imposed on either the wavefunction or the nucleon coordinates and no relative co-ordinates are used. In the *MCRM*, the total angular momentum is the sum of those of the collective rotation and intrinsic motion. The *MCRM* is valid for any nuclear interaction and for a system of fermions or bosons, depending on whether the intrinsic wavefunction is anti-symmetrized or symmetrized respectively.

The *MCRM* equation is also shown to be reducible to that of the phenomenological [34-36,3] and microscopic [37-39] nuclear collective rotation-vibration models, and to the equation of the phenomenological nuclear particle-plus-rotor model [4-7,8,12,28,40] and other microscopic and phenomenological collective models [41-46].

In Section 2, we present the derivation of the *MCRM* Schrodinger equation. In Section 3, we solve the *MCRM* Schrodinger equation. For the model developed in this article, a calculation using a realistic interaction is not necessary because the objective in this article is to compare the predictions of the *MCRM* and *CCRM* and reveal the underlying assumptions and approximation implicit in the *CCRM*, and the areas of differences between the two models. These assumptions, approximations, and differences are present and transparent to various degrees for any nuclear interaction. Therefore, the simplest interaction, namely the self-consistent mean-field deformed harmonic oscillator potential, would be the most transparent and appropriate. Clearly, any discrepancies found between the *MCRM* and *CCRM* would have to be explored in more detail in subsequent analyses using a more realistic interaction. Therefore, in solving the *MCRM* Schrodinger equation, we use a self-consistent single-particle mean-field anisotropic harmonic oscillator potential for the nuclear interaction. In Section 4, we use the *MCRM* to predict the excitation energy and quadrupole moment in the ground-state rotational band of the nucleus $^{20}_{10}Ne$ and compare the results with those of the *CCRM* and with empirical data. In Section 5, we reduce the *MCRM* equation to that of the *CCRM* assuming a small collective-rotation angular momentum and an appropriate microscopic prescription for the combination of the rigid and irrotational flows. In Section 5, we also reduce the *MCRM* equation to the equation of each of the models mentioned in the preceding paragraph. Section 6 presents concluding remarks.

---

[4] In the derivation of the *MCRM*, we try as much as possible not to use the phrase "rotating frame", which is a classical-mechanic concept of a frame rotating with a well defined orientation angle and angular velocity as in the *CCRM*.

## 2. Derivation of microscopic rigid-irrotational flow cranking model

The *MCRM* is derived by transforming the nuclear stationary Schrodinger equation (instead of the Hamiltonian) using the collective rotation-intrinsic product wavefunction for rotation about the *x* or 1 axis, (similar to that in [47])[5]:

$$\Psi = G(\theta) \cdot \Phi(x_{nj}) \tag{9}$$

where $\theta(x_{nj})$ is a collective-rotation angle and is a function of the space-fixed nucleon co-ordinate $x_{nj}$ ($n = 1,..., A;\ j = 1,2,3,$ where $A$ = nuclear mass number) are the space-fixed nucleon co-ordinates. The angle $\theta$ defines the orientation in space of the anisotropic particle distribution (such as quadrupole distribution) described by the intrinsic wavefunction $\Phi$, which is also a function of the space-fixed particle co-ordinates[6]. Applying $\dfrac{\partial}{\partial x_{nj}}$ and $\dfrac{\partial^2}{\partial x_{nj}^2}$ to $\Psi$ in Eq. (9), we obtain:

$$\frac{\partial \Psi}{\partial x_{nj}} = \frac{\partial G}{\partial x_{nj}} \Phi + G \frac{\partial \Phi}{\partial x_{nj}} \tag{10}$$

$$\begin{aligned}
\frac{\partial^2 \Psi}{\partial x_{nj}^2} &= \frac{\partial^2 G}{\partial x_{nj}^2} \Phi + 2 \frac{\partial G}{\partial x_{nj}} \cdot \frac{\partial \Phi}{\partial x_{nj}} + G \frac{\partial^2 \Phi}{\partial x_{nj}^2} \\
&= \Phi \frac{\partial^2 \theta}{\partial x_{nj}^2} \cdot \frac{dG}{d\theta} + \Phi \frac{\partial \theta}{\partial x_{nj}} \cdot \frac{\partial \theta}{\partial x_{nj}} \cdot \frac{d^2 G}{d\theta^2} + 2 \frac{\partial \theta}{\partial x_{nj}} \cdot \frac{dG}{d\theta} \cdot \frac{\partial \Phi}{\partial x_{nj}} + G \frac{\partial^2 \Phi}{\partial x_{nj}^2}
\end{aligned} \tag{11}$$

---

[5] The wavefunction in Eq. (9) is a one-dimensional version of the nuclear-rotor-model wavefunction [4,6,7,12]. Note that $\theta$ depends on the spatial distribution of the nucleons and it is not explicitly a function of the nucleon spin. However, since the nucleon spatial distribution is determined by the intrinsic wavefunction $\Phi$, which depends on the nucleon spins, $\theta$ depends indirectly on the spin. The restriction of the rotation to one spatial dimension is of classical nature but it is adopted from the conventional cranking model because the objective here is to drive a quantum mechanical analogue of the *CCRM*. This classical feature will be removed when the microscopic model is generalized to 3-D rotation.

[6] Note that we do not use any relative co-ordinates for $\Phi$ or anywhere else in the analysis in this article.

Substituting Eq. (11) into the stationary Schrodinger equation:

$$H \cdot \Psi \equiv \left( \frac{1}{2M} \sum_{n,j=1}^{A,3} p_{nj}^2 + V \right) \cdot \Psi = E \cdot \Psi \quad (12)$$

where $M$ is the nucleon mass and $V$ is an arbitrary nuclear interaction, we obtain:

$$G \cdot H \cdot \Phi - \frac{\hbar^2}{M} \cdot \sum_{n,j} \frac{\partial \theta}{\partial x_{nj}} \cdot \frac{\partial G}{\partial \theta} \cdot \frac{\partial \Phi}{\partial x_{nj}} - \frac{\hbar^2}{2M} \cdot \Phi \cdot \sum_{n,j} \frac{\partial \theta}{\partial x_{nj}} \cdot \frac{\partial}{\partial \theta} \left( \frac{\partial \theta}{\partial x_{nj}} \cdot \frac{\partial G}{\partial \theta} \right) = E \cdot \Phi \quad (13)$$

We require the orientation $\theta$ of the nuclear deformed nucleon distribution to be defined by the motion of the particles and hence by the angular momentum operator $L$ along $x$ or 1[7]. Therefore, $\theta$ and $L$ are a canonically conjugate pair, satisfying the commutation relation:

$$[\theta, L] = i\hbar \quad \Rightarrow \quad L \equiv \sum_n \left( y_n p_{nz} - z_n p_{ny} \right) = -i\hbar \frac{\partial}{\partial \theta} \quad (14)$$

Substituting Eq. (14) into Eq. (13), we obtain:

$$G \cdot H \cdot \Phi + \frac{1}{M} \cdot \sum_{n,j} \frac{\partial \theta}{\partial x_{nj}} \cdot (L \cdot G) \cdot p_{nj} \cdot \Phi + \frac{1}{2M} \cdot \Phi \cdot \sum_{n,j} \frac{\partial \theta}{\partial x_{nj}} \cdot L \cdot \left( \frac{\partial \theta}{\partial x_{nj}} \cdot L \cdot G \right) = E \cdot \Phi \quad (15)$$

Next we assume that $G$ is an eigenstate of $L$:

$$L e^{i\gamma\theta} = \hbar \gamma e^{i\gamma\theta} \quad (16)$$

where $\hbar\gamma$ is the angular momentum associated with the collective rotation. $\gamma$ is determined later in this section. Substituting Eq. (16) into Eq. (15), we obtain:

$$H \cdot \Phi + \frac{\hbar\gamma}{M} \cdot \sum_{n,j} \frac{\partial \theta}{\partial x_{nj}} \cdot p_{nj} \cdot \Phi + \frac{\hbar^2 \gamma^2}{2M} \cdot \sum_{n,j} \frac{\partial \theta}{\partial x_{nj}} \cdot \frac{\partial \theta}{\partial x_{nj}} \cdot \Phi - \frac{i\gamma\hbar^2}{2M} \cdot \Phi \cdot \sum_{n,j} \frac{\partial^2 \theta}{\partial x_{nj}^2} = E \cdot \Phi \quad (17)$$

We define the rotation angle $\theta$ in terms of the nuclear quadrupole distribution since observations (experimental and theoretical) indicate that nuclear rotational motion is dominated by the quadrupole nucleon distribution (Bohr-Mottelson's quadrupole deformation model and numerous other collective models such as Villars' collective model using quadrupole moment to define the rotation angle are a testament to this fact). In line with this observation, we define a rotation angle $\theta$ to satisfy the relation (for rotation about $x$ or 1 axis only)[8]:

$$\frac{\partial \theta}{\partial x_{nj}} = \sum_{k=1}^{2} \chi_{jk}\, x_{nk}, \quad \chi_{jk} = 0 \text{ for } j,k \neq 2,3 \tag{18}$$

---

[7] Note that $L$ can be considered to be the total angular momentum including the particle spin because $\theta$ does not depend explicitly on the spin as discussed in footnote 5.

The real 3x3 matrix $\chi$ can be chosen to be a sum of different types of matrices, each describing a different type of physical motion such as quadrupole rigid and irrotational, and non-quadrupole rigid flow regimes described in [41-46,48-50]. In this article, we choose $\chi$ to be the sum of a symmetric and an antisymmetric matrices so that the non-zero elements of $\chi$ are $\chi_{23} \equiv \chi_2 + \chi_3$, and $\chi_{32} = -\chi_2 + \chi_3$. We choose $\chi_3 = \lambda \cdot \chi_2$ and hence:

$$\chi_{23} \equiv (1+\lambda)\cdot \chi_2, \text{ and } \chi_{32} = -(1-\lambda)\cdot \chi_2 \tag{19}$$

Substituting Eq. (19) into $[\theta, L] = i\hbar$ in Eq. (14), we obtain:

$$\chi_2 = -\left(\mathcal{I}_+ - \lambda \cdot \mathcal{I}_-\right)^{-1} \equiv -\mathcal{I}^{-1} \tag{20}$$

where the intrinsic rigid-flow $\mathcal{I}_+$ and deformation $\mathcal{I}_-$ moments of inertia are defined as:

$$\mathcal{I}_+ \equiv \sum_n \left(y_n^2 + z_n^2\right), \quad \mathcal{I}_- \equiv \sum_n \left(y_n^2 - z_n^2\right) \tag{21}$$

We now substitute Eqs. (18)-(20) into Eq. (17), and ignore the fourth term on the left-hand side of Eq. (17) arising from the action of $L$ on $\mathcal{I}_+$ and $\mathcal{I}_-$ (i.e., the term arising from the interaction of rotation with fluctuations in intrinsic nucleon quadrupole distribution) because this term is relatively small, and its expectation over the state $\Phi$ generally vanishes, and such terms are excluded from in the *CCRM* since the angular velocity is a constant in the *CCRM*. Eq. (17) then becomes:

---

[8] Classically, $\dfrac{\partial \theta}{\partial x_{nj}}$ may be considered to be the collective component of the particle velocity field, refer to [47] for more detail. For any linear (in Eq. (18)) or other flow prescription for $\theta$, one can prove (using Eqs. (18)-(20) or Stoke's theorem, refer to [46, Eq. (57)]) that, for a system of more than one particle, the mixed second partial derivatives of $\theta$ are discontinuous, i.e., $\vec{\nabla}_n \times \vec{\nabla}_n \theta \neq 0$. This discontinuity seems to be related to the observation that a change $\delta\theta$ in the collective angle $\theta$ corresponds to different sets of changes $\delta\vec{r}_n$ in the particle positions in a multi-particle system. Even for a single particle, $\vec{\nabla} \times \vec{\nabla} \theta \neq 0$ at the coordinate system origin. However, this discontinuity is of no consequence for the analysis presented in this article because no mixed second derivative of $\theta$ appears anywhere in the analysis and all the derived variables are continuous and well behaved.

$$\left[ H + \frac{\hbar \gamma}{M \mathcal{J}} \cdot (L - \lambda \cdot T) + \frac{\hbar^2 \gamma^2}{2M \mathcal{J}^2} \cdot \left[ \left(1 + \lambda^2\right) \cdot \mathcal{J}_+ - 2\lambda \cdot \mathcal{J}_- \right] \right.$$
$$\left. + \frac{2i\hbar^2 \gamma \lambda^2}{M \mathcal{J}^3} \cdot \left(\lambda \cdot \mathcal{J}_+ - \mathcal{J}_-\right) \cdot \sum_n y_n z_n \right] \cdot \Phi = E \cdot \Phi \quad (22)$$

where $T$ is a linear shear operator, generating a linear irrotational flow, and defined by:

$$T \equiv \sum_n \left( y_n p_{nz} + z_n p_{ny} \right) \quad (23)$$

We note that the order of the appearance of the operators in the second term on the left-hand-side in Eq. (22) is immaterial because we can readily show that:

$$\left[ \mathcal{J}, L - \lambda \cdot T \right] = 0 \quad (24)$$

for any c-number $\lambda$.

The first term on the left-hand-side of Eq. (22) is the intrinsic energy of the interacting nucleons. The second term on the left-hand-side of Eq. (22) is the rigid-irrotational flow cranking or Coriolis energy term, i.e., the energy associated with the interaction of the motion of the nucleons with the collective rotation. The third term on the left-hand-side of Eq. (22) is the kinetic energy of the collective rigid-irrotational rotation of the nucleus as a whole. It may be viewed as a rigid-irrotational centrifugal energy. The remaining term on the left-hand-side of Eq. (22) is the energy associated with the interaction of the collective rotation with fluctuations/vibrations in the rigid-irrotational intrinsic moments of inertia appearing in Eqs. (20) and (21). These remarks and those in [40] may provide a better understanding of the various interactions involved in the rotational motion.

Defining the rigid-plus-shear-flow angular frequency $\omega_{rs}$:

$$\omega_{rs} \equiv \frac{\hbar \gamma}{M \cdot \mathcal{J}}, \quad (25)$$

we express Eq. (22) in the following generalized *CCRM* form:

$$\left\{ H + \omega_{rs} \cdot (L - \lambda \cdot T) + \frac{M \omega_{rs}^2}{2} \cdot \left[ \left(1 + \lambda^2\right) \cdot \mathcal{J}_+ - 2\lambda \cdot \mathcal{J}_- \right] \right.$$
$$\left. + \frac{2i\hbar \omega_{rs} \lambda^2}{\mathcal{J}^2} \cdot \left(\lambda \cdot \mathcal{J}_+ - \mathcal{J}_-\right) \cdot \sum_n y_n z_n \right\} \cdot \Phi = E \cdot \Phi \quad (26)$$

The excitation energy and effective dynamic moment of inertia are defined in Eqs. (7) and (8). The variable $\mathcal{J}$ in Eqs. (25) and (26) may be called the kinematic moment of inertia associated with the Coriolis energy term, i.e. the inertial mass associated with the interaction of the motion of the nucleons with the collective rotation. This moment of inertia differs from the kinematic moment of inertia $\left[(1+\lambda^2)\cdot\mathcal{I}_+ - 2\lambda\cdot\mathcal{I}_-\right] = 2\mathcal{I} - (1-\lambda^2)\cdot\mathcal{I}_+$ in Eq. (26) that is associated with the kinetic energy of the collective rotation of the nucleus as a whole, i.e., with the centrifugal energy. The Coriolis energy term modifies the intrinsic system properties such as its natural frequencies. On the other hand, the centrifugal energy term contributes more to the collective-rotation excitation energy than does the Coriolis energy term. In contrast, in the *CCRM* (Eq. (6)), the Coriolis energy term affects both the intrinsic system properties and excitation energy.

We determine the collective-rotation angular momentum $\hbar\gamma$ in Eq. (25) by requiring the expectation of the total angular momentum operator $L$ with respect to the wavefunction $\Psi$ in Eq. (9) to have the experimentally observed rotational-band excited-state angular momentum $\hbar J$ [9]:

$$\hbar J = \langle \Psi | L | \Psi \rangle = \langle G | L | G \rangle + \langle \Phi | L | \Phi \rangle = \hbar\gamma + \hbar l = \omega_{rs} \cdot M \cdot \mathcal{I} + \hbar l \quad (27)$$

where:

$$\hbar l \equiv \langle \Phi | L | \Phi \rangle \quad (28)$$

The prescription in Eq. (27) differs from that in Eq. (5) for the *CCRM* by the collective angular momentum $\hbar\gamma$, and reduces to it for small $\hbar\gamma$.

It follows from Eq. (27) that, in the *MCRM*, $\gamma$ (and hence $\omega_{rs}$) and $l$ can have different signs, and hence the collective and intrinsic system rotations can be in the opposite directions, unlike that in the *CCRM* but similarly to that in the particle-plus-rotor model [4-7,8,12,28,40].

The above-stated differences between the *MCRM* and *CCRM* generate some significant differences in the predictions of the two models as is demonstrated in Section 4.

For $\lambda = 0$ (i.e., for the rigid collective flow only), Eqs. (25) to (27) reduce respectively to:

$$\omega_{rig} \equiv \frac{\hbar\gamma}{M\cdot\mathcal{I}_+}, \quad (29)$$

$$\left(H + \omega_{rig}\cdot L + \frac{1}{2}\omega_{rig}^2\cdot M\cdot\mathcal{I}_+\right)\cdot\Phi = E\cdot\Phi \quad (30)$$

---

[9] The value of $\gamma$ determined by Eq. (27) may be viewed as an approximation (implied by the *CCRM*) to an integer value of $\gamma$ needed to ensure that $\Psi$ is single-valued function of $\theta$.

$$\hbar J = \langle \Psi | L | \Psi \rangle = \langle G | L | G \rangle + \langle \Phi | L | \Phi \rangle = \hbar \gamma + \hbar l = \omega_{rig} \cdot M \cdot \mathcal{I}_+ + \hbar l \qquad (31)$$

where $\omega_{rig}$ is the rigid-flow angular frequency and $M \cdot \mathcal{I}_+$ is the rigid-flow moment of inertia (defined in Eq. (21)) and it commutes with $L$. Eq. (30) resembles the *CCRM* equation in a space-fixed frame given in Eq. (6), but differs from Eq. (6) by the rigid-flow centrifugal energy term (third term on the left-hand-side of Eq. (30)), and by the microscopically defined $\omega_{rig}$ instead of the constant parameter $\omega_{cr}$ in Eq. (6). Eq. (30) also differs from the *CCRM* Eq. (4) in the rotating frame by the aforementioned terms and by the sign of $\omega_{rig}$ (note that the *MCRM* solves Eq. (30) whereas the *CCRM* solves Eq. (4) and not Eq. (6)). Eq. (30) is time-reversal invariant whereas Eq. (6) is not.

3. **Solutions of Eqs. (4), (26), and (30)**

To quantitatively compare the predictions of the *MCRM* and *CCRM*, we must solve Eqs. (4), (26), and (30) for a given nuclear interaction $V$ in Eq. (12). In this article, we use the simple deformed harmonic oscillator potential:

$$H = \frac{1}{2M} \cdot \sum_{n,j=1}^{A,3} p_{nj}^2 + \frac{M\omega_1^2}{2} \cdot \sum_n x_n^2 + \frac{M\omega_2^2}{2} \cdot \sum_n y_n^2 + \frac{M\omega_3^2}{2} \cdot \sum_n z_n^2 \qquad (32)$$

for $V$ because it yields analytical solutions and hence facilitates identification and resolution of any differences between the predictions of the *MCRM* and *CCRM*. Of-course, subsequent calculations using realistic $V$ need to be performed to realistically quantify the impact of the discrepancies. Since in the *CCRM* angular velocity $\omega_{cr}$ is a constant (and hence the *CCRM* ignores fluctuations in the moments of inertia), to compare the *MCRM* and *CCRM* predictions we suppress fluctuations in the moments $\mathcal{I}_+$ and $\mathcal{I}_-$, and hence in $\mathcal{J}$ and $\omega_{rs}$ by replacing $\mathcal{I}_+$ and $\mathcal{I}_-$ by their expectation values over the intrinsic state $\Phi$ (noting that the expectation of the fourth term on the left-hand-side of Eq. (26) vanishes)[10]. Defining these expectations by:

$$\mathcal{I}_+^o \equiv \langle \Phi | \mathcal{I}_+ | \Phi \rangle, \quad \mathcal{I}_-^o \equiv \langle \Phi | \mathcal{I}_- | \Phi \rangle, \quad \mathcal{J}^o \equiv \langle \Phi | \mathcal{J} | \Phi \rangle, \quad \omega_{rs}^o \equiv \frac{\hbar \gamma}{M \mathcal{J}^o} \qquad (33)$$

we then solve Eq. (26) using a method similar to those in references [51-55] used to solve the *CCRM* Eq. (4), namely using a canonical or unitary transformation to eliminate the cross terms $y_n p_{nz}$ and $z_n p_{ny}$ in Eq. (26), and transform the Hamiltonian $H + \omega_{rs} \cdot (L - \lambda \cdot T)$ in Eq. (26) to the harmonic oscillator Hamiltonian:

---

[10] We may account for these fluctuations and their interaction with the collective rotation by regarding $\mathcal{J}$ as an independent variable and transform Eq. (26) to it using the method given in [39,56,57].

$$\bar{H} = \frac{1}{2M} \cdot \sum_{n,j=1}^{A,3} p_{nj}^2 + \frac{M\omega_1^2}{2} \cdot \sum_n x_n^2 + \frac{M\alpha_2^2}{2} \cdot \sum_n y_n^2 + \frac{M\alpha_3^2}{2} \cdot \sum_n z_n^2 \tag{34}$$

and obtain the energy eigenvalue in a space-fixed frame:

$$E = \hbar\omega_1 \Sigma_1 + \hbar\alpha_2 \Sigma_2 + \hbar\alpha_3 \Sigma_3 + \frac{1}{2}\omega_{rs}^{o\,2} \cdot M \cdot \left[ \left(1+\lambda^2\right) \cdot \mathcal{I}_+^o - 2\lambda \cdot \mathcal{I}_-^o \right] \tag{35}$$

where:

$$\alpha_2^2 \equiv \omega_+^2 + (1-\lambda^2) \cdot \omega_{rs}^{o\,2} + \sqrt{\omega_-^4 + 4\omega_{rs}^{o\,2} \cdot \omega_\lambda^2}, \quad \alpha_3^2 \equiv \omega_+^2 + (1-\lambda^2) \cdot \omega_{rs}^{o\,2} - \sqrt{\omega_-^4 + 4\omega_{rs}^{o\,2} \cdot \omega_\lambda^2} \tag{36}$$

$$\omega_+^2 \equiv \frac{\omega_2^2 + \omega_3^2}{2}, \quad \omega_-^2 \equiv \frac{\omega_2^2 - \omega_3^2}{2}, \quad \Sigma_k \equiv \sum_{n_k=0}^{n_{kf}} (n_k + 1/2), \quad \omega_\lambda^2 \equiv \omega_+^2 + \lambda \cdot \omega_-^2 \tag{37}$$

where $n_{kf}$ is the number of oscillator quanta in the $k^{\text{th}}$ direction at the Fermi surface.

For $\bar{H}$ in Eq. (34) to approximate a Hartree-Fock mean-field Hamiltonian, we minimize the energy $E$ in Eq. (35) with respect to the frequencies $\omega_k$ ($k = 1, 2, 3$) at a fixed value of $J$ and hence of $\mathcal{I}^o$, $\omega_{rs}^o$, $\mathcal{I}_+^o$, and $\mathcal{I}_-^o$ given by the constraint in Eq. (27), subject to the constant nuclear-quadrupole-volume condition:

$$\langle x^2 \rangle \cdot \langle y^2 \rangle \cdot \langle z^2 \rangle = c_o \tag{38}$$

where $\langle x_k^2 \rangle \equiv \langle \Phi | \sum_n x_{nk}^2 | \Phi \rangle$ ($k = 1, 2, 3$) and $c_o$ is a constant. This minimization yields a self-consistency between the shapes of nuclear equi-potential and equi-density surfaces [7,51-54]. The minimization is performed numerically as in [53,54].

The solution in Eqs. (35), (36), and (37) is specialized to that for Eq. (30) by setting $\lambda$ to zero, and to the solution of Eq. (4) by setting $\lambda$ to zero and $\omega_{rs}^o$ to $-\omega_{cr}$.

We may determine the arbitrary parameter $\lambda$ in Eq. (35) by minimizing the energy $E$ in Eq. (35) with respect to $\lambda$. A study of the results of the above derivations and the calculation in Section 4 shows that most terms in $\lambda$ in the derived expressions cancel one another and the remaining terms in $\lambda$ are negligibly small. Therefore, the contribution to the energy $E$ in Eq. (35) from the operator $\lambda \cdot T$ in Eq. (26) is small and the term $\lambda \cdot T$ in Eq. (26) can be ignored. Therefore, the minimization of the energy $E$ in Eq. (35) with respect to $\lambda$ yields:

$$\lambda = \frac{\mathcal{J}_-^o}{\mathcal{J}_+^o} \tag{39}$$

Substituting Eq. (39) into Eq. (35), we obtain:

$$E = \frac{\omega_{rs}^{o\,2} \cdot M}{2\mathcal{J}_+^o} \cdot \left[\mathcal{J}_+^{o2} - \mathcal{J}_-^{o2}\right] + \hbar\omega_1 \Sigma_1 + \hbar\alpha_2 \Sigma_2 + \hbar\alpha_3 \Sigma_3 \tag{40}$$

Now $\mathcal{J}_-^o$ decreases with $J$ and vanishes at some $J$ depending on parameters $\omega_{rs}^o$ and $\alpha_k$. When this happens, the first term on the right-hand-side of Eq. (40) reduces to the rigid-flow centrifugal kinetic energy.

## 4. Model predictions for $^{20}_{10}Ne$

In this section we compare quantitatively the *MCRM* and *CCRM* predictions for $^{20}_{10}Ne$ ground-state rotational band using the solutions given in Section 3 for the simplest possible interaction, namely the deformed harmonic oscillator potential. This potential is used because it gives an analytic solution and hence facilitates the identification and explanation of the source of the differences between the predictions of the two models. A more realistic potential will be used later to realistically quantify the impact of these differences. The model results are also compared with the measured data. For $^{20}_{10}Ne$, we use the anisotropic-harmonic-oscillator nucleon-occupation configuration $(\Sigma_1, \Sigma_2, \Sigma_3) = (14, 14, 22)$, with the spherical harmonic oscillator frequency $\hbar\omega_o = 35.4 \cdot A^{-1/3}$ $MeV$ as in [53,54].

First we use the rigid-flow *MCRM*, for which $\lambda = 0$ and which resembles most closely the *CCRM*, using various versions of the self-consistency conditions to learn how the model behaves. Fig 1 compares the excitation energy $\Delta E_J$ measured and predicted by the *MCRM* and *CCRM* when the $\omega_k$'s are kept constant at their ground-state (i.e., $J = 0$) self-consistent values (i.e., the nucleus remains prolate at all $J$ values). Fig 1 shows that the *MCRM* predicts well the excitation energy and the *CCRM* predicts a slightly lower $\Delta E_J$. The measured and predicted excitation energies increase with $J$, except at $J = 8$ where the measured $\Delta E_J$ is significantly lower than that predicted by either of the models. This difference may be attributed to a Coriolis-force induced partial alignment with the rotation axis of the quasi-particle angular momenta, which effect is not included in the models. We note that the rotational band predicted by either the *MCRM* or *CCRM* does not terminate at $J = 8$ as expected because no nuclear shape transition occurs for the particular self-consistency used.

Fig 2 shows that the measured quadrupole moment $Q_o$ decreases monotonically with $J$ and there is a sharp drop in $Q_o$ at $J = 8$ when the nucleus presumably becomes symmetric about the rotation axis (i.e., when $\omega_2$ and $\omega_3$ become equal). The *MCRM* and *CCRM* predict lower $Q_o$, which decreases gradually up to $J = 4$ in the *MCRM* and up to $J = 8$ in the *CCRM*, and increases above these $J$ values (the increase in $Q_o$ for the *CCRM* is not included in Fig 2).

The intrinsic angular momentum predicted by the *MCRM* decreases from about $-6.5\hbar$ at $J = 2$ to $-8\hbar$ at $J = 8$, and in this range of $J$, the collective angular momentum increases from $8.5\hbar$ to $16\hbar$. The predicted rotational band terminates above $J = 23$ when the intrinsic oscillator frequency $\alpha_3$ vanishes and hence the model governing equations have no solutions. The results in Fig 2 indicate that, to predict the observed change in the shape of the nucleus (i.e., $Q_o$) the *MCRM* and *CCRM* must account for nuclear volume conservation at all $J$ values as is normally done.

Figs 3 and 4 show respectively the excitation energies and quadrupole moment when we use the $\omega_k$ values determined self-consistently at all $J$ values. The *CCRM* predicts higher $\Delta E_J$ above $J = 4$, and a $Q_o$ that decreases similarly to the observed $Q_o$ up to $J = 6$, above which point the nucleus becomes progressively axially symmetric about the rotation axis (i.e., $\omega_2$ and $\omega_3$ become equal) and the rotational band terminates at $J = 8$. The *MCRM* predicts significantly higher $\Delta E_J$, and a $Q_o$ that decreases rapidly up to $J = 4$ and increases slightly thereafter. The predicted intrinsic angular momentum decreases from $-6.5\hbar$ at $J = 2$ to $-8\hbar$ at $J \geq 4$, and the collective angular momentum is $8.5\hbar$ at $J = 2$ and $12\hbar, 14\hbar, 16\hbar$ at $J = 4, 6, 8$ respectively.

The results presented above show that the rigid-flow *MCRM* predicts somewhat similar results to those of the *CCRM*. However, the collective and intrinsic-system rotations are in opposite directions, and the intrinsic system angular momentum reaches its limiting value of $-8\hbar$ at $J = 2$ and remains at this value thereafter while the collective angular momentum continues to increase with $J$, unlike the situation in the *CCRM*, where the intrinsic angular momentum is equal to $J$ at all values of the rotating-frame angular velocity.

For a non-zero $\lambda$, i.e., for rigid-plus-**irrotational flow** *MCRM*, the results are similar to those of the rigid-flow *MCRM*.

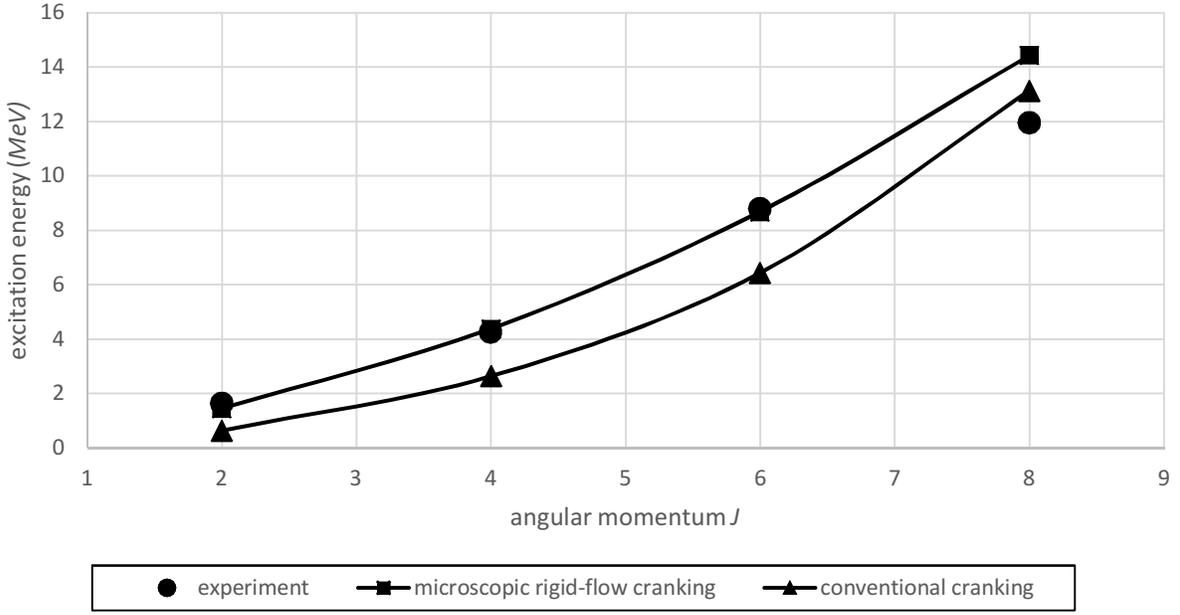

Fig 1: excitation energy versus J for self-consistency at J = 0 only

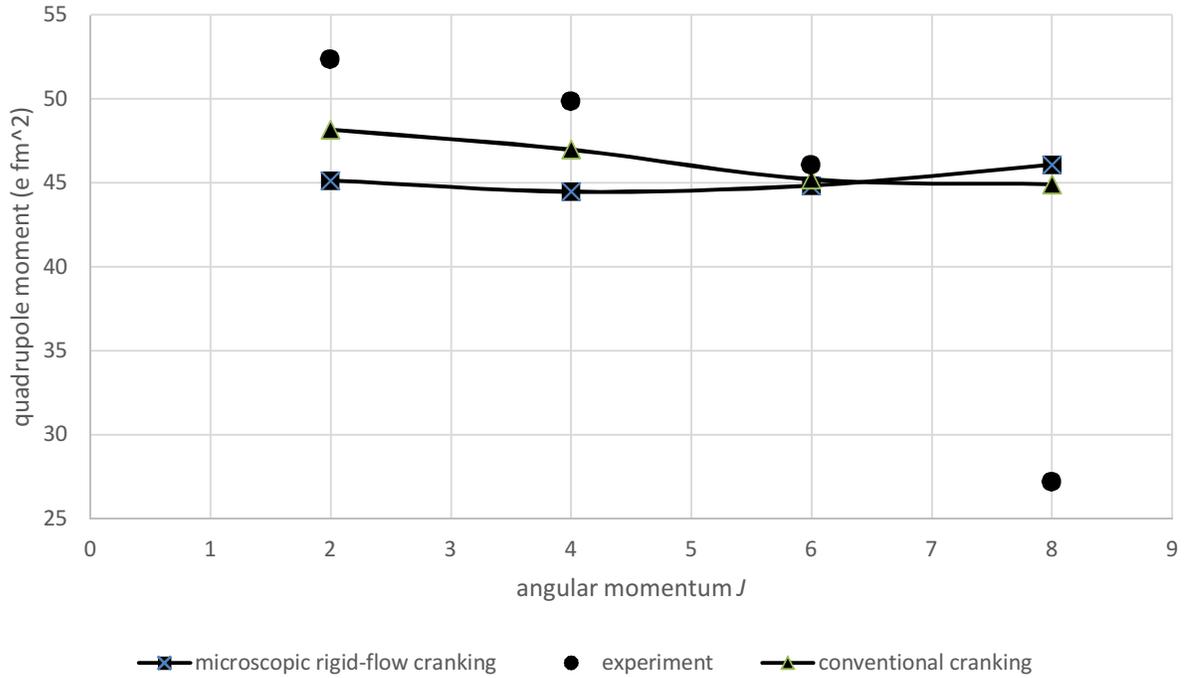

Fig 2: quadrupole moment versus J for self-consistency at J = 0 only

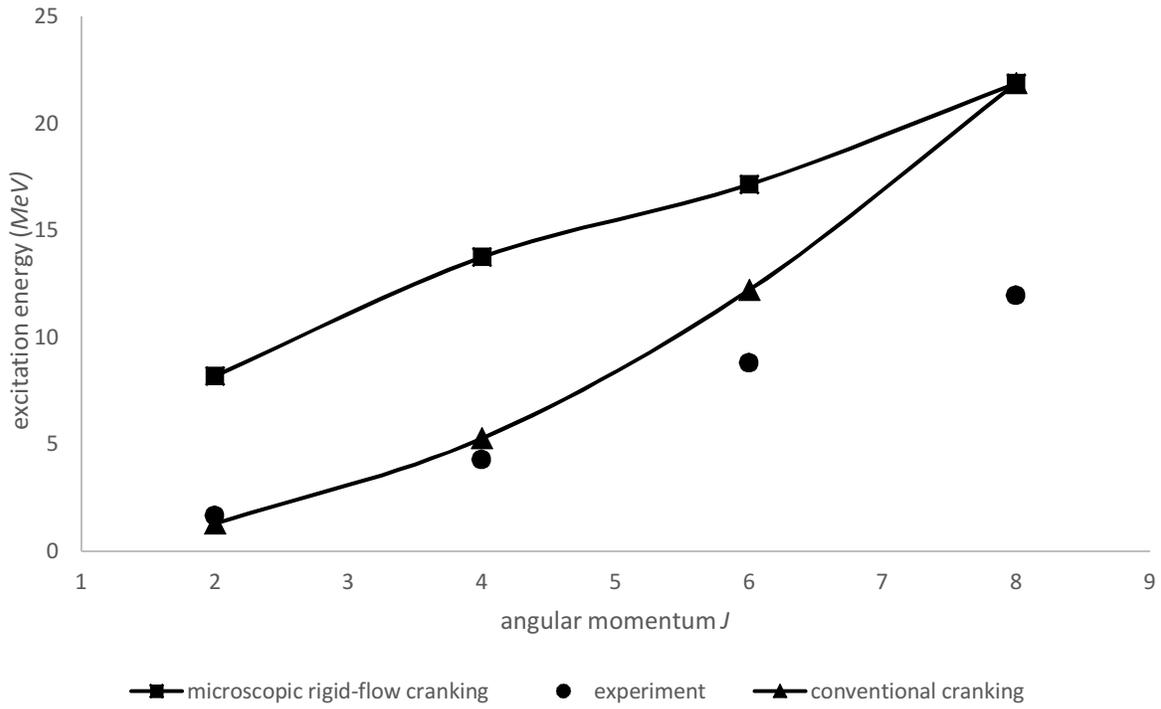

Fig 3: excitation energy versus J for self-consistency at all J

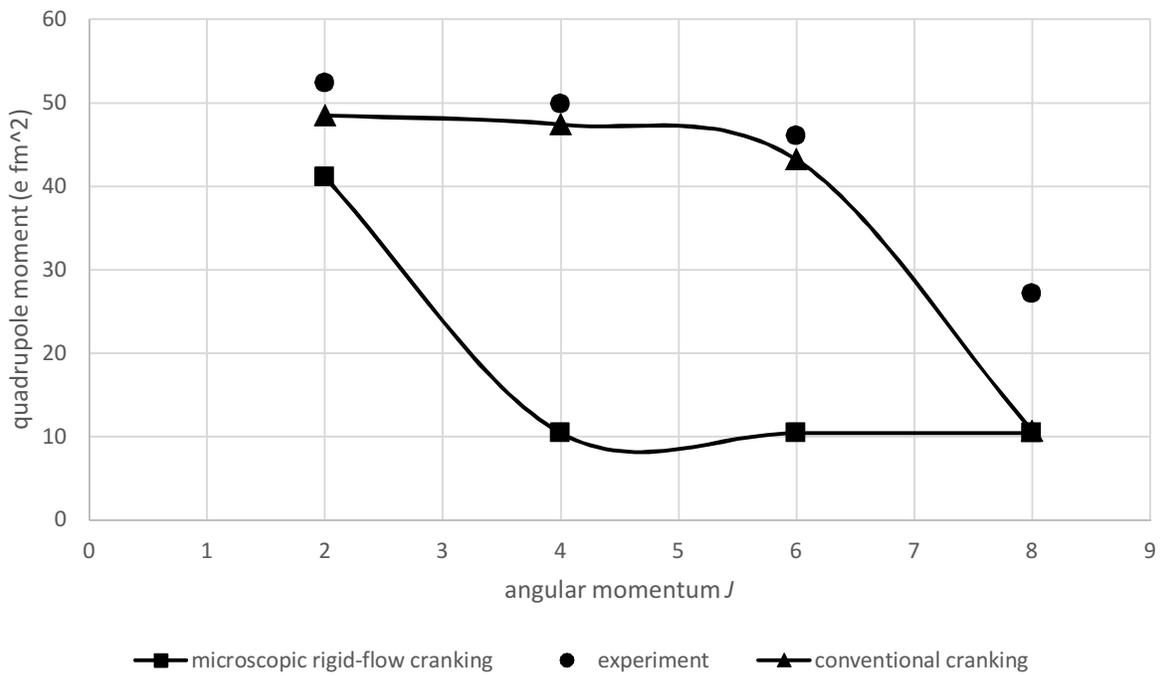

Fig 4: quadrupole moment versus J for self-consistency at all J

## 5. Correspondence between microscopic cranking and other collective rotational models

In this section, we reduce that the *MCRM* Eq. (26) to the equation of the *CCRM*, nuclear particle-plus-rotor model, phenomenological and microscopic nuclear collective rotation-vibration models, and phenomenological classical collective rotation models.

### 5.1 Correspondence with conventional cranking model

The results of the calculation in Section 4 indicate that the *MCRM* predicts values of $\omega_{rs}^o$ significantly higher than $\omega_{cr}$ predicted by the *CCRM*. This happens because we have determined $\omega_{rs}^o$ using $\hbar\gamma = \omega_{rs}^o \cdot M \cdot \mathcal{J}^o$ (Eq. (33)) and the constraint $J = \gamma + l$ (Eq. (31)). On the other hand, the *CCRM* uses the constraint $J = l$, and hence assumes implicitly that the angular momentum $\hbar\gamma$ of the rotating frame is negligibly small. To match the *MCRM* and *CCRM* equations, we must therefore assume that $\gamma$ is negligibly small. Eq. (31) then becomes:

$$l = J - \gamma \approx J \qquad (41)$$

Furthermore, the parameters in the *MCRM* Eq. (26) are nearly independent of $\lambda$, i.e., the contribution from the term $\lambda \cdot T$ in Eq. (26) is small, as noted in Section 3 in connection with Eq. (39). Therefore, we can drop the term involving $\lambda$ in $l$. Eq. (41) then becomes identical to Eq. (5), and hence $\omega_{rs}^o$ becomes identical to $-\omega_{cr}$. It then follows that the *MCRM* energy $E$ in Eq. (35) and the *CCRM* energy $E_{cr}$ in Eq. (6) become identical if the centrifugal energy in Eq. (35) is set equal to the Coriolis energy term in Eq. (6), i.e., if the following equation holds:

$$\frac{1}{2}\omega_{cr} \cdot M \cdot \left[(1+\lambda^2) \cdot \mathcal{J}_+^o - 2\lambda \cdot \mathcal{J}_-^o \right] = \langle \Phi_{cr} | L | \Phi_{cr} \rangle \equiv \hbar l \qquad (42)$$

where we have set $\omega_{rs}^o = -\omega_{cr}$. Solving Eq. (42) for $\lambda$, we obtain:

$$\lambda = \frac{\mathcal{J}_-^o}{\mathcal{J}_+^o}\left(1 + \sqrt{1 + \frac{2\mathcal{J}_+^o \cdot \hbar l}{\omega_{cr} \cdot \mathcal{J}_-^{o2}} - \frac{\mathcal{J}_+^{o2}}{\mathcal{J}_-^{o2}}}\right) \qquad (43)$$

where all the parameters on the right-hand-side of Eq. (43) are nearly independent of $\lambda$. We note that the factor $\dfrac{\mathcal{J}_-^o}{\mathcal{J}_+^o}$ minimizes the energy $E$ in Eq. (35) as stated in connection with Eq. (39).

By construction, the excitation energies predicted by the *MCRM* and *CCRM* are identical. However, Fig 6 shows that the value of $\gamma$ predicted by the *MCRM* Eq. (33), i.e., $\hbar\gamma = \omega_{rs}^o \cdot M \cdot \mathcal{J}^o$, is not small at all and in fact increases with $J$ because both $\mathcal{J}^o$ and $\omega_{rs}^o$ increase with $J$ as Fig 6 shows.

The inconsistency noted in the preceding paragraph between the value of $\gamma$ assumed (i.e., zero, as implied by the *CCRM*) and predicted (from $\hbar\gamma = \omega_{rs}^o \cdot M \cdot \mathcal{J}^o$) by the *MCRM* may be resolved by adding other flow regimes (such non-quadrupole rigid flow) to the rigid-irrotational flow regime in Eqs. (18) and (19) such that, in the prescription $\hbar\gamma = \omega_{rs}^o \cdot M \cdot \mathcal{J}^o$, $\mathcal{J}^o$ decreases as $\gamma$ decreases but $\omega_{rs}^o$ remains finite (which seems to be implied in the comparison between the *CCRM* and rotor models in [8], refer to Section 5.2 for more discussion). This approach will be considered in a future article.

## 5.2 Correspondence with particle-plus-rotor model

As noted in Section 5.1, we can drop the small term $\lambda \cdot T$ in Eq. (26) to obtain:

$$\left\{ H + \omega_{rs}^o \cdot L + \frac{\hbar^2 \gamma^2}{2M\mathcal{J}^{o2}} \left[ \left(1 + \lambda^2\right) \cdot \mathcal{J}_+^o - 2\lambda \cdot \mathcal{J}_-^o \right] \right\} \cdot \Phi = E \cdot \Phi \qquad (44)$$

When $\omega_{rs}^o$ is replaced by $\dfrac{\hbar\gamma}{M \cdot \mathcal{J}^o}$ (Eq. (33)) and $\gamma$ is replaced by $J - l$ (Eq. (27)), Eq. (44) becomes:

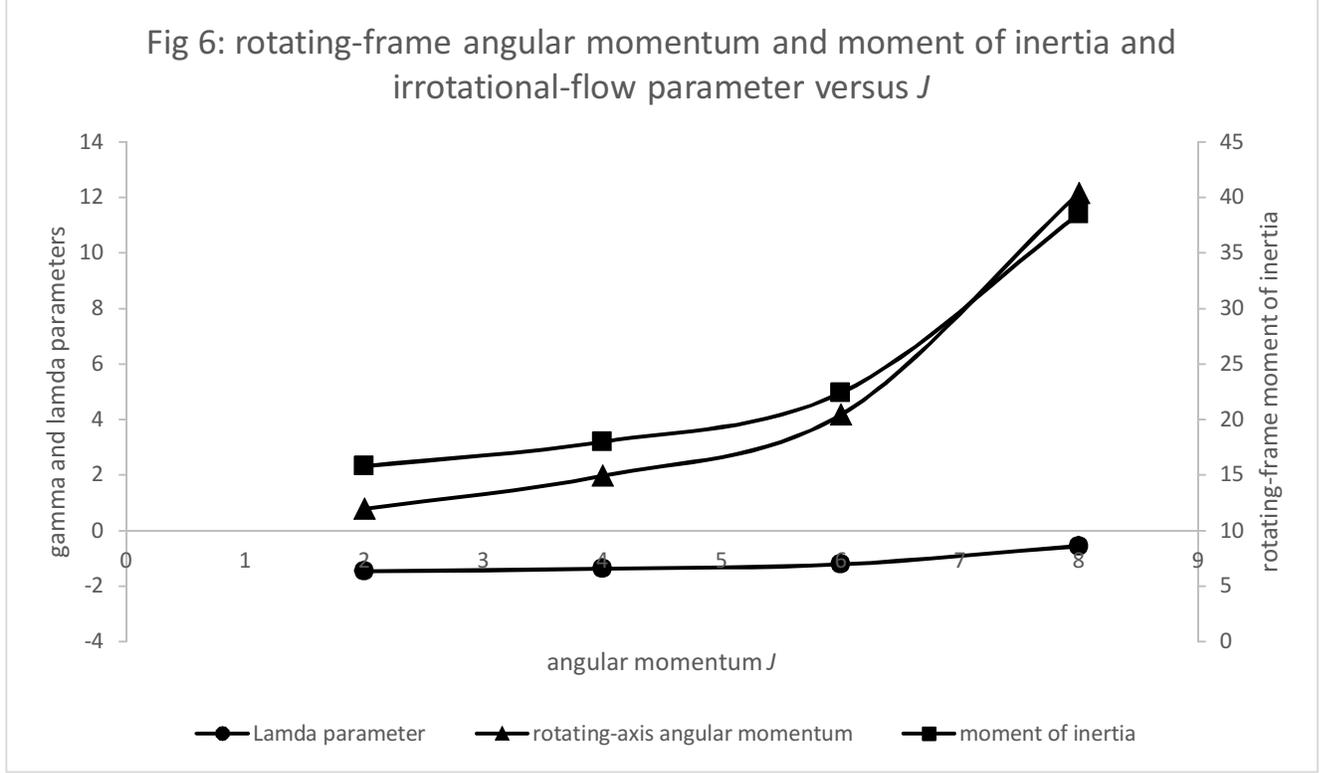

Fig 6: rotating-frame angular momentum and moment of inertia and irrotational-flow parameter versus J

$$\left\{ H + \frac{\hbar(J-l)}{M \cdot \mathcal{I}^o} \cdot L + \frac{\hbar^2 (J-l)^2}{2M \mathcal{I}^{o2}} \left[ (1+\lambda^2) \cdot \mathcal{I}_+^o - 2\lambda \cdot \mathcal{I}_-^o \right] \right\} \cdot \Phi = E \cdot \Phi \qquad (45)$$

The Hamiltonian of the nuclear particle-plus-rotor model for rotation about a single axis is [4-7,12,28,40,58]:

$$H_{rp} \equiv H + \frac{\hbar (J-\hat{j})^2}{\mathcal{I}_{rp}} \qquad (46)$$

where $\hat{j}$ is the sum of the angular momentum operators of all the valence (out-of-core) nucleons and $\mathcal{I}_{rp}$ is the core moment of inertia. The mean-field part of $H_{rp}$ in Eq. (46) is:

$$H_{rp} \equiv H + \frac{2\hbar(J-j)}{\mathcal{I}_{rp}} \cdot (J - \hat{j}) = H - \frac{2\hbar(J-j)}{\mathcal{I}_{rp}} \cdot \hat{j} + \frac{2\hbar(J-j)J}{\mathcal{I}_{rp}} \qquad (47)$$

where $j$ is the mean of $\hat{j}$. We may identify the second and the third terms on the right-hand-side of Eq. (47) with respectively the second and third terms on the left-hand-side of Eq. (45), keeping in mind that the angular momentum operator $L$ includes the angular momenta of all the nucleons whereas $\hat{j}$ includes only the valence (out of core) nucleons. This comparisons shows the correspondence between the microscopic cranking and rotor-plus-particle models.

The correspondence between the *MCRM* and *CCRM* and between the *MCRM* and particle-plus-rotor model establishes a correspondence between the *CCRM* and particle-plus-rotor model. The latter correspondence elucidates the comparison between the rotor model and *CCRM* presented in [8]. We note that in the rotor model in [8], the only coupling between the rotational and intrinsic motion is through the assumed three moments of inertia, which are calculated using the *CCRM*. We therefore, infer, from a comparison of Eq. (46) and the rotor model equation in [8], that the rotor model assumes the condition $L \cdot \Phi = 0$. However, in [8] a quantum analogue of the *CCRM* was surmised in the following form (for rotation along a single axis):

$$H_{rp} \equiv H - \frac{\hbar J \cdot L}{\mathcal{J}_{rp}} \tag{48}$$

Eq. (48) resembles the coupled $J \cdot L$ term in the first term on the left-hand-side of Eq. (45).

### 5.3 Correspondence with phenomenological and microscopic rotational models

When we impose zero angular momentum constraint (i.e., $L \cdot \Phi = 0$) on the intrinsic wavefunction $\Phi$ so that $\gamma = J$, and set $\lambda = 0$, Eq. (44) becomes:

$$\left( H + \frac{\hbar^2 J^2}{2M \mathcal{J}_+^o} \right) \cdot \Phi = E \cdot \Phi \tag{49}$$

Eq. (49) is identical to the rotational part in the equations of the microscopic [39,56,57] and phenomenological [3] collective rotation-vibration models for rotation about a single axis.

### 5.4 Correspondence with phenomenological classical collective rotational models

When we impose the constraints $L \cdot \Phi = 0$ and $H \cdot \Phi = 0$ (i.e., no intrinsic motion) so that $\gamma = J$, Eq. (44) becomes:

$$\frac{\hbar^2 J^2}{2M \mathcal{J}^{o2}} \left[ (1 + \lambda^2) \cdot \mathcal{J}_+^o - 2\lambda \cdot \mathcal{J}_-^o \right] \cdot \Phi = E \cdot \Phi \tag{50}$$

Eq. (50) may considered to be a quantum version of the classical collective two-fluid rotational models [41,42,43,44] with only one (namely $\lambda$) free parameter (the other parameter is determined in Eq. (20) by the commutation relation $[\theta, L] = i\hbar$).

### 6. Concluding remarks

The conventional cranking model (*CCRM*) is often used to study rotational properties of nuclei. In view of its importance in nuclear structure studies and its phenomenological and semi-classical basis, it is desirable to have a better understanding of the assumptions and approximations that underlie the model derivation. In the hope of achieving this objective, we have attempted, in this article, to derive the model simply and from first principles by transforming the nuclear Schrodinger equation (instead of the Hamiltonian) using a collective rotation-intrinsic product wavefunction and imposing no constraints on either the wavefunction or the particle co-ordinates and using only space-fixed nucleon co-ordinates. The collective-rotation angle is chosen canonically conjugate to the total angular momentum, and defined by a combination of rigid and irrotational collective flows of the nucleons, instead of being imposed externally as in the *CCRM*. As a consequence, the microscopic cranking model (*MCRM*) Schrodinger equation is time reversal invariant, unlike the *CCRM*.

The resulting *MCRM* Schrodinger equation for the intrinsic wavefunction resembles that of the *CCRM* in a space-fixed frame. However, the *MCRM* equation has, in addition to the rigid-irrotational flow Coriolis energy term, a rigid-irrotational flow centrifugal kinetic energy term, and a rotation-intrinsic fluctuation interaction energy term that are absent from the *CCRM* equation. Furthermore, the angular velocity in the *MCRM* is not a constant parameter but is a dynamical variable determined by the collective-rotation angular momentum and an associated kinematic moment of inertia. This moment is determined by the nature of the aforementioned rigid-irrotational flow combination. The collective-rotation angular momentum is determined by equating the expectation of the total angular momentum to the angular momentum of a rotational-band state. This expectation is a sum of the angular momenta of the collective rotation and intrinsic system. It turns out that the collective and intrinsic rotations are in the opposite directions.

We compare quantitatively the *MCRM* and *CCRM* predictions for the simplest possible interaction, namely a deformed harmonic oscillator potential. This potential is used because it yields an analytic solution to the model equations and hence facilitates the identification and resolution of the source of any differences between the predictions of the two models. A more realistic potential will be used later to quantify the impact of these differences. For this comparison, we suppress the fluctuations in the kinematic moments of inertia and angular velocity by replacing the moments by their expectation values in each intrinsic-system state because the *CCRM* ignores such fluctuations. The oscillator frequencies are determined self-consistently from numerical minimization of the energy subject to a constant nuclear volume condition as in the *CCRM*.

Using the *MCRM*, we calculate the excitation energy and quadrupole moment for the ground-state rotational band of the nucleus $^{20}_{10}Ne$. The results for rigid-flow *MCRM* (i.e., for the irrotational-flow component ignored) show that the model predictions have trends similar to those of the *CCRM* except that the intrinsic angular momentum has a significantly higher maximum absolute value (in order to reduce the high value of the collective angular momentum to the rotational-band state angular momentum). The predicted excitation energy is significantly higher than the measured one. The predicted quadrupole moment is smaller than that measured. Adding the irrotational-flow component to the rigid-flow *MCRM* is found to somewhat improve the agreement with the results of the *CCRM* and measurement at $J = 2$. However, at higher $J$ values the irrotational-flow component has little effect because the collective-rotation moment of inertia is reduced to the rigid-flow moment.

It is shown that the *MCRM* equation becomes identical to those of the *CCRM* for a particular choice of the combination of the rigid and irrotational flows and when the collective-rotation angular momentum is assumed to be negligibly small. However, even though these results may indicate that the *CCRM* implicitly assumes that the rotating-frame has negligibly small angular momentum and that the frame rotation is to some extent governed by combined rigid and irrotational flows of the nucleons, the collective-rotation angular momentum predicted by *MCRM* is not small and increases with $J$. In future studies, we will attempt to resolve these inconsistencies between the *MCRM* and *CCRM* by including other types of collective flows in the *MCRM* to reduce or eliminate the rotating- frame angular momentum. We also intend to generalize the *MCRM* to three spatial dimensions to, among other things, remove the non-quantum feature of rotation about a single axis.

We have also shown that the *MCRM* equation can be reduced to that of the nuclear particle-plus-rotor model, microscopic and phenomenological collective rotation-vibration models, and phenomenological classical collective rotational models.